\documentclass[12pt]{article}
\usepackage[mathscr]{eucal}
\usepackage{amssymb,latexsym}
\usepackage{verbatim}
\usepackage{amsmath}
\usepackage{amsthm}
\usepackage{enumerate}

\newtheorem{thm}{Theorem}[section]
\newtheorem{lem}[thm]{Lemma}
\newtheorem{Def}{Definition}[section]
\newtheorem{cor}[thm]{Corollary}

\newcommand\calV{{\mathcal{V}}}

\newcommand\calT{{\mathcal{T}}}

\newcommand\calH{{\mathcal{H}}}

\newcommand\Th{\Theta}

\renewcommand\l{\lambda}

\newcommand\bbR{{\mathbb R}}

\renewcommand\S{\Sigma}

\renewcommand\d{\partial}

\newcommand\f{\phi}
\renewcommand\L{\triangle}
\newcommand\D{\nabla}
\newcommand\e{\epsilon}

\renewcommand\div{{\rm div}}

\newcommand\la{\langle}
\newcommand\ra{\rangle}

\renewcommand\l{\lambda}

\newcommand\8{\infty}

\renewcommand\th{\theta}
\renewcommand\Th{\Theta}

\newcommand\<{\la}
\renewcommand\>{\ra}

\newcommand\beq{\begin{equation}}
\newcommand\eeq{\end{equation}}
\newcommand\ben{\begin{enumerate}}
\newcommand\een{\end{enumerate}}
\newcommand\bit{\begin{itemize}}
\newcommand\eit{\end{itemize}}

\newcounter{mnotecount}[section]

\title{Rigidity of marginally trapped surfaces\\ and the topology of black holes}

\author{Gregory J. Galloway
 \\Department of Mathematics \\ University of Miami
}

\begin{document}
\date{}
\maketitle
\vspace{.2in}

\begin{abstract} 
In a recent paper \cite{GS}  the author and Rick Schoen
obtained a generalization to higher dimensions of a  classical result of Hawking concerning the topology of black holes.  It was proved  that, apart from certain exceptional
circumstances,  cross sections of the event horizon, in the
stationary case,  and `weakly outermost' marginally outer
trapped surfaces, in the general case,
in black hole spacetimes obeying the dominant energy condition, are of positive Yamabe type.  This implies many well-known restrictions on the topology,
and is consistent with recent examples  of five
dimensional stationary black hole spacetimes with horizon topology
$S^2 \times S^1$.  In the present paper,  
we rule out for `outermost' marginally outer
trapped surfaces, in particular, for cross sections of the event
horizon in stationary black hole spacetimes,
the possibility of any such exceptional circumstances (which might have permitted,
e.g., toroidal cross sections).  This follows from the main result,
which is a rigidity result for marginally outer
trapped surfaces that are not of positive Yamabe type.
\end{abstract}


\section{Introduction}  

Some recent developments in physics inspired by string theory, such as the AdS/CFT 
correspondence and brane world phenomenology,
have heightened interest in higher dimensional gravity.  In particular, 
there has been a considerable amount of recent research devoted to the study of
black holes in higher dimensions; for a sample, see \cite{ER,peet,cvetic},
and references cited therein. 
In \cite{GS},  Schoen and the author obtained
a generalization to higher dimensions of a classical result of Hawking concerning the topology of black
holes.  We proved that, apart from certain exceptional circumstances,
`weakly outermost' marginally outer trapped surfaces, in particular cross sections of the event horizon
in stationary black hole spacetimes, are of positive Yamabe type, i.e., admit
metrics of positive scalar curvature, provided the dominant energy condition holds.
This implies many well-known restrictions on the topology of the horizon,
and is consistent with recent examples \cite{ER} of five
dimensional stationary black hole spacetimes with horizon topology
$S^2 \times S^1$.  In particular, in $3+1$
dimensions, the Gauss-Bonnet theorem implies that the horizon is topologically a $2$-sphere, and one recovers  Hawking's theorem.  

If, however, certain quantities vanish on the horizon, e.g., if the horizon is Ricci flat and  spacetime is vacuum in its vicinity,  then the arguments in \cite{GS} do not quite
guarantee the conclusion of being positive Yamabe.   One of the main aims of the
present paper is to rule out the possibility of any exceptions to being positive Yamabe under a natural set
of physical circumstances.  This will follow as a consequence of a rigidity
result for marginally outer trapped surfaces that do not admit metrics of positive scalar
curvature. This result may be viewed as a spacetime analogue of the rigidity results for  area minimizing hypersurfaces in a Riemannian manifold obtained in \cite{cai, CG}.  The rationale for such a result had been discussed by the author (in the $3+1$ setting)
in \cite{G}.

Before stating our main results, let us begin with a few definitions,
and, in particular, introduce the basic object of study, that of a marginally
outer trapped surface.  Let $\S^{n-1}$, $n \ge 3$, be a compact spacelike
submanifold of co-dimension two in a spacetime (time-oriented Lorentzian manifold) $(M^{n+1}, g)$.
Under suitable orientation assumptions, $\S$ admits two smooth nonvanishing
future directed null normal vector fields $K_+$ and $K_-$.  These vector fields
are unique up to pointwise scaling.  By convention, we refer to $K_+$ as outward pointing  and $K_-$ as inward pointing.  Let $\chi_{\pm}$ denote the null second fundamemtal form associated to $K_{\pm}$.  Thus, for each $p \in \S$, 
$\chi_{\pm} : T_p\S \times T_p\S \to \bbR$ is the symmetric bilinear form defined
by,
\begin{align}
\chi_{\pm}(X, Y) = \<\D_X K_{\pm}, Y\> \quad \mbox{for all } X,Y \in T_p\S \,,
\end{align}
where $\D$ is the Levi-Civita connection of $g = \<\,,\,\>$.   
Tracing with respect
to the induced metric $h$ on $\S$ we obtain the null expansion scalars
(or null mean curvatures) $\th_{\pm} ={\rm tr}\, \chi_{\pm} = \div_{\S}K_{\pm}$.
As is well-known, the sign of $\th_{\pm}$ is invariant under positive rescalings
of $K_{\pm}$.  Physically, $\th_+$ (resp., $\th_-$) measures the divergence
of the  outward pointing (resp., inward point) light rays emanating
from $\S$.  
For round spheres in Euclidean slices of Minkowski space,
with the obvious choice of inside and outside, one has $\th_- < 0$ and $\th_+ >0$.
In fact, this is the case in general for large ``radial" spheres in asymptotically flat spacelike hypersurfaces.    However, in regions of spacetime where the gravitational field
is strong, one may have both $\th_- < 0$ and $\th_+ < 0$, in which case $\S$
is called a trapped surface.  Under appropriate energy and causality conditions,
the occurrence of a trapped surface signals  the onset of gravitational collapse
\cite{P} and the existence of a black hole \cite{HE}.  

Focusing attention on just the outward null normal, we say that $\S$ is 
an {\it outer trapped surface} if $\th_+ < 0$, and is a {\it marginally outer trapped
surface} (MOTS) if $\th_ + = 0$.    MOTSs arise in a number of natural situations.
For example, compact cross sections of the event horizon in stationary
(steady state) black
hole spacetimes are MOTSs.  For dynamical black hole spacetimes, 
MOTSs typically occur in the black hole region, i.e., the region inside the event horizon.  
While there are heuristic arguments for the existence of MOTSs in this situation, 
based on looking at the boundary of the `trapped region' \cite{HE,W} within a given
spacelike slice, a recent
result of Schoen \cite{S, AM2, E} rigorously establishes their existence under natural
conditions.   MOTSs are the key ingredient behind 
the development of quasi-local notions of black holes
(see \cite{AK} and references cited therein).
On the more purely mathematical side, 
there are connections between MOTSs in spacetime and minimal
surfaces in Riemannian manifolds.   In fact, 
a MOTS contained in a 
{\it totally geodesic} spacelike hypersurface $V^n \subset M^{n+1}$ is 
simply a minimal hypersurface in $V$.   Despite the absence of a
variational characterization of MOTSs like that for minimal surfaces, 
MOTS have recently been  shown to satisfy a number of analogous properties; see, e.g.,
\cite{AG, AMS, AMS2, AM, AM2, E, GS, S}, as well as the important earlier work of Schoen and
Yau~\cite{SY2}. 
The rigidity results presented here provide another case in point. 

For our main results, we shall only consider spacetimes $(M^{n+1},g)$ that
satisfy the Einstein equations,
\begin{align}
R_{ab} - \frac12 R g_{ab} = T_{ab} 
\end{align}
for which the energy-momentum tensor $\calT$ obeys the dominant
energy condition,  $\calT(X,Y) = T_{ab}X^aY^b \ge 0$ for all future pointing
causal vectors $X,Y$.

We now restrict attention to MOTSs contained in a spacelike hypersurface.  
Thus, let $V^n$ be an $n$-dimensional, $n \ge 3$, spacelike
hypersurface in a spacetime $(M^{n+1},g)$, and let $\S^{n-1}$ be a closed
hypersurface in $V^n$. Assume that $\S^{n-1}$ separates
$V^n$ into an ``inside" and an ``outside".  Denote the closure of the
outside of $V$ by $V_+$; hence $V_+$ is a manifold with boundary $\d V_+ = \S$.

We adopt the following terminology.  
\begin{Def}\label{outermost}
Let $\S^{n-1}$ be a MOTS in a spacelike hypersurface $V^n$, as above.
\ben[(i)]
\item We say that $\S$ is an outermost MOTS in $V$ provided there are no
outer trapped or marginally outer trapped surfaces outside of,
and homologous to,~$\S$.
\item We say that $\S$ is a weakly outermost MOTS in $V$ provided there are no
outer trapped  surfaces outside of,
and homologous to, $\S$.
\een
\end{Def}

\smallskip
\noindent
{\it Remarks:}
\ben[(1)]
\item We note that $\S$ is an outermost MOTS if and only if there
are no {\it weakly outer trapped} surfaces  ($\th_+ \le 0$) outside of, and
homologous to, $\S$.  The point is, if $S$ is  weakly outer trapped then either
it's a MOTS or else it can be perturbed, via null mean curvature flow,
to an outer trapped surface \cite[Lemma 5.2]{AM2}. 
\item By the existence result of Schoen alluded to above \cite{S, AM2, E}, under a natural outer barrier condition (which always holds in the asymptotically flat case), and provided the dimension is not too high, there exists outside of each outer trapped surface a MOTS homologous to it.  Hence, under these circumstances, an outermost MOTS, as defined here, is outermost in the conventional sense.  
\item Heuristically, a weakly outermost MOTS $\S$ is the ``outer limit" of outer trapped surfaces in $V$.  Weakly outermost MOTSs were referred to as outer apparent horizons
in \cite{GS}. 
\een

One of the main aims of this paper is to present a proof of the following theorem.

\begin{thm}\label{main}
Let $(M^{n+1},g)$, $n\ge 3$, be a spacetime satisfying the dominant energy condition,
and let $\S^{n-1}$ be an outermost MOTS in  a spacelike hypersurface $V^n$.
Then $\S^{n-1}$ is of positive Yamabe type, i.e., admits a metric of positive
scalar curvature.  
\end{thm}

In fact, we shall prove the following rigidity result,   which immediately implies
Theorem \ref{main}. 

\begin{thm}\label{rigid}
Let $(M^{n+1},g)$, $n\ge 3$, be a spacetime satisfying the dominant energy condition,
and let $\S^{n-1}$ be a weakly outermost MOTS in  a spacelike hypersurface $V^n$.
If $\S^{n-1}$ does not admit a metric of positive scalar curvature then there exists
a neighborhood $U \approx [0,\e) \times \S$ of $\S$ in $V_+$ such that each slice
$\S_t = \{t\} \times \S$, $t \in [0,\e)$ is a MOTS.  In fact each  such slice
has vanishing outward null second fundamental form,  $\chi_+ = 0$, and is Ricci
flat.  
\end{thm}


It is also shown that a  certain energy-momentum term vanishes along each slice.
Theorem \ref{rigid} shall be proved in two stages.  The first stage, and the
main effort of the paper is to prove Theorem \ref{rigid} subject to the
additional assumption that $V^n$ has {\it nonpositive} mean curvature,
$\tau \le 0$;\footnote{By our sign conventions, the hyperbola
$t = - \sqrt{1+x^2}$ in Minkowski $2$-space has negative mean curvature.} 
see Theorem \ref{rigid2} in Section 3. The second stage uses  a ``deformation"
argument to derive Theorem \ref{rigid} from Theorem \ref{rigid2}.   
While Theorem \ref{rigid2} is a pure ``initial data" result, the proof of Theorem
\ref{rigid} makes use of the enveloping spacetime.  Theorem~\ref{main} shows
that, for outermost MOTS, the exceptional case in the main result of \cite{GS}
can be eliminated.

A basic fact about standard $(3+1)$-dimensional black hole spacetimes \cite{HE, W} obeying the
null energy condition is that there can be
no outer trapped, or even marginally outer trapped, surfaces outside the event
horizon.   The proof, which relies on the Raychaudhuri equation \cite{HE,W}, also works in higher
dimensions.   Thus, Theorem \ref{rigid} implies the following.

 
 \begin{cor}\label{posyam2}
 Cross sections \footnote{By  {\it cross section}, we  mean
smooth compact intersection of the event horizon with a spacelike hypersurface.
}
 of the event horizon in stationary black hole
spacetimes obeying the dominant energy condition are of positive 
Yamabe type.  
 \end{cor}
 
 In particular, there can be no toroidal horizons.
 The proof of Theorem \ref{rigid} is presented in Section 3, following
some preliminary results, presented in Section 2.

\section{Analytic and geometric preliminaries}
Let $(\S,h)$ be a compact Riemannian manifold.  
We draw together here various facts (all essentially known) about  operators $L : C^{\infty}(\S) \to C^{\infty}(\S)$ of the form
\begin{align}\label{op}
L(\f) = - \triangle \f   + 2\<X,\D \f\> + (Q + \div X - |X|^2) \f  \,,
\end{align}
where $Q \in C^{\infty}(\S)$, $X$ is a smooth vector field on $\S$ and 
$\<\,,\,\> = h$.  The stability operator associated with variations in the
null expansion, as explicitly introduced in \cite{AMS}, is of this form.  

As discussed in \cite{AMS}, although $L$ is not self-adjoint in general, the Krein-Rutman theorem, together with other arguments,  implies the following.

\begin{lem}\label{prin}  Let $\l_1 = \l_1(L)$ be the principal eigenvalue of $L$ (eigenvalue
with smallest real part).   Then the following hold.
\ben[(i)]
\item  $\l_1$ is real and simple.  There
exists an associated eigenfunction $\f$  ($L(\f) = \l_1 \f$) which is strictly positive.
\item $\l_1 \ge 0$ (resp., $\l_1 > 0$) if and only if there exists $\psi \in C^{\infty}(\S)$, $\psi> 0$, such that $L(\psi) \ge 0$ (resp., $L(\psi) > 0$). 
\een
\end{lem}

We wish to compare $L$ with the ``symmetrized" operator $L_0 : C^{\infty}(\S) \to C^{\infty}(\S)$, obtained  by setting $X=0$,
\begin{align}\label{op2}
L_0(\f) = - \triangle \f + Q\, \f  \,.
\end{align}

The main argument in  \cite{GS} shows that if $\l_1(L) \ge 0$ then $\l_1(L_0) \ge 0$.
In fact, as noted by Mars and Simon \cite{MS}, a simple tweaking of this 
argument gives the following.  
\begin{lem}\label{compare} 
The principal eigenvalues $\l_1(L)$ of $L$ and $\l_1(L_0)$ of $L_0$
satisfy, $\l_1(L) \le \l_1(L_0)$.
\end{lem}
\proof In inequality (2.7) in \cite{GS}, replace ``$\ge 0$" by ``$= \l_1\, \f$", and proceed.
\qed

\medskip
A key result in the Schoen-Yau study of manifolds of positive scalar curvature~\cite{SY}
is that a compact stable minimal hypersurface in a manifold of positive scalar curvature
admits, itself,  a metric of positive scalar curvature.  Related results have been
obtained in~\cite{ACG,GS}, and are proved using a simplification of the original argument of Schoen and Yau due to Cai \cite{cai}.  These results may be formulated in a slightly more
general context, as follows. 

\begin{lem}\label{scalar} 
Consider the operator $L_0 = -\L + Q$ on $(\S, h)$, with 
\begin{align}\label{Q}
Q = \frac12 S - P \,,
\end{align}
where $S$ is the scalar curvature of $(\S,h)$ and $P \ge 0$. 
If $\l_1(L_0) \ge 0$ then $\S$ admits a metric of positive scalar curvature,
unless $\l_1(L_0) = 0$, $P \equiv 0$ and $(\S,h)$ is Ricci flat.  

\end{lem}
\proof 
Let  $\phi \in C^{\8}(\S)$ be a positive eigenfunction associated to the eigenvalue
$\l_1 = \l_1(L_0)$.
The scalar curvature $\tilde S$ of $\S$  
in the conformally rescaled metric $\tilde h = \f^{\frac2{n-2}}h$ is then given by,
\begin{align}\label{rescale}
\tilde S & =  \phi^{-\frac{n}{n-2}}(-2\L\f + S \f +\frac{n-1}{n-2}\frac{|\D\f|^2}{\f}) \nonumber \\\  
& =  \phi^{-\frac{2}{n-2}}(2\l_1 + 2P +\frac{n-1}{n-2}\frac{|\D\f|^2}{\f^2})
\end{align} 
where the second equation follows from (\ref{op2}), (\ref{Q}) and the fact
that $L_0(\f) = \l_1 \f$.  Since all terms
in the parentheses above are nonnegative, (\ref{rescale}) implies that $\tilde S \ge 0$. 
If $\tilde S > 0$ at some point, then by well known results
\cite{KW} one can conformally rescale $\tilde h$  to a metric 
of strictly positive scalar curvature.  If, on the other hand, $\tilde S$ vanishes
identically, then (\ref{rescale}) implies: $\l_1 = 0$, $P \equiv 0$ and $\f$ is constant.  Equations (\ref{op2})  and (\ref{Q}) then  imply that $S \equiv  0$.  By an 
argument of Bourguignon (see \cite{KW}),  one can then deform $h$  in the direction
of the Ricci tensor of $\S$ to obtain a metric of positive scalar curvature, unless  $(\S,h)$ is Ricci flat.   ~\qed

Finally, Lemmas (\ref{compare}) and (\ref{scalar}) combine to give the following.
\begin{lem}\label{scalar2}
Lemma \ref{scalar} also holds for the operator $L$ in (\ref{op}), with 
$Q$ as in (\ref{Q}).
\end{lem}

Apart from the conclusion that $\l_1(L) = 0$ (if $\S$ does not admit
a metric of positive scalar curvature), this was proved, in a specific
context, in \cite{GS}. 

\section{Proof of Theorem \ref{rigid}} 

Let the notation and terminology be as in the statement of Theorem \ref{rigid},
and the discussion leading up to it.  As discussed in the introduction, we begin
by proving Theorem \ref{rigid}, subject to a restriction on the mean curvature
of $V^n$. 
\begin{thm}\label{rigid2}
Let $(M^{n+1},g)$, $n\ge 3$, be a spacetime satisfying the dominant energy condition,
and let  $V^n$ be  a spacelike hypersurface in $M^{n+1}$ with mean curvature
$\tau \le 0$. Suppose $\S^{n-1}$ is a weakly outermost MOTS in  $V^n$
that does not admit a metric of positive scalar curvature.  Then there exists
a neighborhood $U \approx [0,\e) \times \S$ of $\S$ in $V_+$ such that each slice
$\S_t = \{t\} \times \S$, $t \in [0,\e)$ is a MOTS.  In fact each  such slice
has vanishing outward null second fundamental form,  $\chi_+ = 0$, and is Ricci flat.
\end{thm}

\proof
The first step is to show that a neighborhood of $\S$ in $V_+$ is foliated 
by {\it constant} null expansion hypersurfaces,  with respect to a suitable scaling of the
future directed outward null normals.

Let $t \to \S_t$ be a variation of  $\S = \S_0$, $-\e <t < \e$, with variation vector field 
$\calV = \left . \frac{\d}{\d t}\right |_{t=0} = \phi \nu$,  $\phi \in C^{\infty}(\S)$, where
$\nu$ is the outward unit normal of $\S$ in $V$.   
Let $\th(t)$ denote
the null expansion of $\S_t$ with respect to $K_t = Z + \nu_t$, where $Z$ is the future directed timelike unit normal to $V$ and $\nu_t$ is the
outer unit normal  to $\S_t$ in $V$.   A computation shows \cite{AMS, AMS2, AM, CG},
\begin{align}\label{der}
\left . \frac{\d\th}{\d t} \right |_{t=0}  & =   L(\f) 
  = -\triangle \phi + 2\<X,\D\phi\>  + \left(Q+{\rm div}\, X - |X|^2 \right)\phi \,,
\end{align}
where,
\beq\label{Q2}
Q = \frac12 S - \calT(Z,K) - \frac12 |\chi|^2   \,,
\eeq
$S$ is the scalar curvature of $\S$, $\chi$ is the null second fundamental
form of $\S$ with respect to $K = \nu +Z$, $X$ is the vector field on $\S$
defined by $X = {\rm tan}\,(\D_{\nu}Z)$, and $\<\,,\,\>$ now denotes
the induced metric  on $\S$.

Let $\l_1$ be the principal eigenvalue of $L$. As per Lemma \ref{prin},
$\l_1$ is real, and there is an associated eigenfunction $\phi$ that
is strictly positive.   Using $\phi$ to
define our variation, we have from (\ref{der}),
\begin{align}\label{der2}
\left . \frac{\d\th}{\d t} \right |_{t=0} =\l_1 \phi \,.
\end{align}
The eigenvalue $\l_1$ cannot be negative, for otherwise (\ref{der2}) would
imply that  $\frac{\d\th}{\d t}< 0$ on $\S$.  Since $\th = 0$ on $\S$, this would mean
that for $t>0$ sufficiently small, $\S_t$ would be outer trapped, contrary 
to assumption.
Thus, $\l_1 \ge 0$, and since  $\S$ does not carry a metric of positive scalar curvature,
we may apply  Lemma \ref{scalar2} to $L$ in (\ref{der}), with 
$P =  \calT(Z,K) +  \frac12 |\chi|^2  \ge 0$, to conclude that $\l_1 = 0$ (and 
also that $Q=0$).

For $u\in C^{\infty}(\S)$, $u$ small, let $\Th(u)$ denote the null  expansion of the hypersurface $\S_u : x \to exp_x u(x) \nu$ with respect to the (suitably normalized)
future directed outward null normal field to $\S_u$.  $\Th$ has linearization, $\Th'(0) = L$.
We introduce the
operator,
\begin{align}
\Th^* : C^{\infty}(\S) \times \bbR \to  C^{\infty}(\S) \times \bbR  \,, \quad 
\Th^*(u,k) = \left(\Th(u) -k , \int_{\S}u\right)  \,.
\end{align}
Since, by Lemma \ref{prin}.1, $\l_1=0$ is a simple eigenvalue, the 
kernel of $\Th'(0)=L$ consists only of constant multiples of the 
eigenfunction $\phi$.   We note that $\l_1=0$ is also a simple eigenvalue
for the adjoint $L^*$ of $L$ (with respect to the standard
$L^2$ inner product on $\S$), for which there exists
a positive eigenfunction $\phi^*$.  Then the equation
$Lu =f$ is solvable if and only if $\int f\phi^* = 0$.
From these facts it follows easily that  
 $\Th^*$ has invertible  linearization about $(0,0)$.  
Thus, by the inverse function theorem,
for $s\in \bbR$ sufficiently small there exists $u(s) \in C^{\8}(\S)$ and $k(s)\in \bbR$ 
such that,
\begin{align}\label{inverse}
\Th(u(s)) = k(s) \qquad \mbox{and} \qquad \int_{\S} u(s) dA = s\,.
\end{align}
By the chain rule, $\Th'(0)(u'(0)) = L(u'(0)) = k'(0)$.    The fact that $k'(0)$ is orthogonal
to $\phi^*$ implies  that $k'(0) = 0$.
Hence $u'(0) \in {\rm ker}\, \Th'(0)$.  The second equation in (\ref{inverse})
then implies that $u'(0) = const \cdot \f > 0$.   

It follows that 
for $s$ sufficiently small, the hypersurfaces $\S_{u_{s}}$ form a smooth foliation
of a neighborhood of $\S$ in $V$ by hypersurfaces of constant null expansion.
Thus, one can introduce coordinates $(t, x^i)$ in a neighborhood $W$ of $\S$ in $V$,
such that, with respect to these coordinates, $W = (-t_0,t_0) \times \S$, and for each
$t \in  (-t_0,t_0)$, the $t$- slice $\S_t = \{t\} \times \S$ has constant null expansion 
$\th(t)$ with respect to $K|_{\S_t}$, where $K = Z+\nu$,
and $\nu$ is the outward unit normal
field to the $\S_t$'s in $V$.  In addition, the coordinates $(t,x^i)$ can be chosen
so that  $\frac{\d}{\d t} = \phi \nu$, 
for some positive function $\phi = \phi(t,x^i)$ on $W$.

A computation  similar to that leading to (\ref{der}) (but where we can no longer
assume $\th$ vanishes)
shows that the null expansion function $\th = \th(t)$ of the foliation obeys the evolution equation, 
\footnote{Although we have checked this independently, Equation \eqref{evolve}
follows easily from Lemma 3.1 in~\cite{AMS2}; see also \cite{AM}.}
\begin{align}\label{evolve}
\frac{d\th}{dt}  = \tilde L_t (\phi)
\end{align}
where, for each $t \in  (-t_0,t_0)$, $\tilde L_t$ is the operator on $\S_t$
acting on $\phi$ according to,
\begin{align}\label{evo-op}
\tilde L_t (\phi) & =  -\triangle \phi + 2\<X,\D\phi\> + \nonumber \\
& \qquad\qquad \left(\frac12 S - \calT(Z,K)   + \th\tau
-\frac12 \th^2 - \frac12 |\chi|^2 
+{\rm div}\, X - |X|^2 \right)\phi \,.
\end{align}
It is to be understood that, for each $t$, the above terms  live on $\S_t$,  
e.g., $\triangle = \triangle_t$ is the Laplacian on $\S_t$, 
$S =S_t$ is the scalar curvature of $\S_t$, and so on. 

The assumption that $\S$ is weakly outermost, together with the constancy of $\th(t)$, implies that $\th(t) \ge 0$ for all $t \in [0,t_0)$.  
Hence, since $\th(0) = 0$, to show that $\th(t) = 0$ for all $t \in [0,t_0)$.
it is sufficient to show that $\th'(t) \le 0$ for 
all $t \in [0,t_0)$.  Suppose there exists $t \in (0,t_0)$ such that
$\th'(t) > 0$.  For this value of $t$,  (\ref{evolve}) implies
$\tilde L_t(\f) > 0$.  Then Lemma~(\ref{prin}) implies that $\l_1(\tilde L_t) > 0$. 
Recalling the assumption $\tau \le 0$,  we may apply  Lemma (\ref{scalar2}) to 
$\tilde L_t$, with $P =  \calT(Z,K)   -  \th\tau + \frac12 \th^2 + \frac12 |\chi|^2 \ge 0$, 
to conclude that $\S_{t} \approx \S$ carries a metric of positive scalar curvature, contrary to assumption. 

Thus, $\th(t) = 0$ for all $t \in [0,t_0)$.  Since, by \eqref{evolve},  $\tilde L_t(\phi) = \th' = 0$, Lemma \ref{prin} implies  $\l_1(\tilde L_t) \ge 0$ for each $t \in [0,t_0)$. Hence, by Lemma (\ref{scalar2}), we have that   for each $t \in [0,t_0)$, $\chi_t  = 0$, $\S_t$ is Ricci flat and  $\calT(Z,K)$ vanishes along $\S_t$.   \qed

\bigskip
\noindent {\it Proof of Theorem \ref{rigid}.} We now show how Theorem~\ref{rigid}
can be obtained from Theorem~\ref{rigid2}. 

Let the setting be as in the statement of Theorem \ref{rigid}.   It is straight
forward to construct a spacelike hypersurface $\tilde V^n$ in $M^{n+1}$ with
the following properties: (i) $\tilde V$ and $V$ meet tangentially along $\S$,
(ii) $\tilde V$ is in the causal past of $V$ and (iii) $\tilde V$ has mean
curvature $\tilde \tau \le 0$.  ($\tilde V$ can be constructed from spacelike curves
orthogonal to  $\S$ and tangent to $V$ at $\S$, having sufficiently large 
curvature, and bending towards the past.)

The condition that $\S$ is weakly outermost in $V$ transfers {\it to a sufficient extent}
to~$\tilde V$, as described in the following claim.

\medskip
\noindent
{\bf Claim.} For every variation $t \to \S_t$, $-\e <t < \e$, of $\S= \S_0$ in $\tilde V$, with variation vector field $\calV = 
\phi\tilde\nu$, 
$\phi > 0$, there exists $t_0 \in (0,\e)$ such that $\S_t$ is not outer trapped
for all $t \in (0,t_0)$.

\proof[Proof of the claim] Suppose, to the contrary, there exists
a variation $t \to \S_t$, $0 \le t < \e$, of $\S$ in $\tilde V_+$ (the outside of $\tilde V$)
and a sequence $t_n \searrow 0$ such that $\S_n := \S_{t_n}$ is outer
trapped.  Let $\calH_n$ be the null hypersurface generated by the
future directed outward null geodesics orthogonal to $\S_n$.  Restricting
to a small tubular neighborhood of $\S$, 
for all $n$ sufficiently large, $\calH_n$ will be a smooth null hypersurface
that meets $V$ in a compact surface $\hat \S_n$ outside of, and homologous
to, $\S$.   By Raychaudhuri's equation
for a null geodesic congruence \cite{HE, W} and the null energy condition (which is
a consequence of the dominant energy condition), the  expansion
of the null generators of $\calH_n$ must be nonincreasing to the future.
It follows that, for $n$ large, $\hat \S_n$ is outer trapped, contrary
to the assumption that $\S$ is weakly outermost. \qed 

\smallskip
Hence, $\S$ is weakly outermost in $\tilde V$, in the restricted sense
of the claim.  But this version of weakly outermost is clearly sufficient for the proof of Theorem \ref{rigid2}.   Thus, by this slight modification of Theorem \ref{rigid2},  there
exists a foliation $\{\tilde\S_u\}$, $0 \le u \le u_0$, of a neighborhood $\tilde U$ of
$\S$ in $\tilde V_+$ by MOTS, $\tilde\th_+(u) =~0$.  Pushing each $\tilde\S_u$ along its
future directed outward null normal geodesics into $V$, we obtain, by taking
$u_0$ smaller if necessary, a smooth foliation $\{\S_u\}$, $0 \le u \le u_0$, of
a neighborhood $U$ of $\S$ in $V_+$.   Moreover, the argument based
on Raychaudhuri's equation used in the claim now implies
that, for each $u \in (0,u_0)$, $\S_u$ is weakly outer trapped, i.e., has null expansion
$\th_+(u) \le 0$. If $\th_+(u) <  0$ at some point, one could perturb $\S_u$ within $V$  to obtain a strictly outer trapped surface in $V$ homologous
to $\S$ (see the first remark after Definition \ref{outermost}).  It follows that each $\S_u$ in the foliation is a MOTS.  Moreover, the same argument as that used
at the end of the proof of Theorem \ref{rigid2} implies that    for each $u \in [0,u_0)$, $\chi_u = 0$, $\S_u$ is Ricci flat and  $\calT(Z,K)$ vanishes along $\S_u$. 
\qed

\medskip

We remark in closing that the curvature estimates of Andersson and Metzger~\cite{AM} provide criteria for extending the local foliation by MOTS  in Theorem~\ref{rigid2} to a global one.

\section*{Acknowledgement} The author would like to thank  
Lars Andersson, Abhay Ashtekar, Robert Bartnik, Mingliang Cai, Piotr Chru\'sciel,
Jan Metzger, Dan Pollack and Walter Simon for useful 
comments and discussions.  This work was supported in part by NSF grants 
DMS-0405906 and DMS-0708048.

\providecommand{\bysame}{\leavevmode\hbox to3em{\hrulefill}\thinspace}

\end{document}